\documentclass[aps,prd]{revtex4}
\usepackage{graphics}
\begin{document} 

\title{
Radiation back reaction on moving branes} 
\author{Ian G. Moss}
\email{ian.moss@ncl.ac.uk}
\author{James P Norman}
\email{j.p.norman@ncl.ac.uk}
\affiliation{School of Mathematics and Statistics, University of  
Newcastle Upon Tyne, NE1 7RU, UK}

\date{\today}

%%%%%%%%%%%%%%%%%%%%%%%%%%%%%%%%%%%%%%%%%%%%%%

\begin{abstract}
This paper addresses the radiation back reaction problem for cosmological
branes. A general framework is provided in which results are given for the
radiation reaction with massles and massive scalar fields with flat extra
dimensions and massless conformal fields in anti-de Sitter extra dimensions.
For massless scalar field radiation the back reaction terms in the equation of
motion are non-analytic. The interpretation of the radiation reaction terms is
discussed and the equations of motion solved in simple cases. Nucleosynthesis
bounds on dark radiation give  a lower bound on the string vacuum energy
scale of $\sqrt{A_T}\,m_p$, where $A_T$ is the tensor perturbation amplitude in
the cosmic microwave background.

\end{abstract}

\maketitle
%%%%%%%%%%%%%%%%%%%%%%%%%%%%%%%%%%%%%%%%%%%
\section{introduction}

Attempts to construct a unified theory of gravity and the other fundamental
forces using superstrings have provided a new impetus for the study of higher
dimensions and branes. In certain low energy limits, superstring theory
reduces to a brane universe, or universes, where ordinary matter is confined
to surfaces embedded in higher dimensions \cite{horava96-2}. The cosmological
evolution of these models can be different from the standard cosmological
scenario \cite{cline99,binetruy99,shiromizu99} and their study is therefore
worthwhile.

The matter fields on the brane interact with higher dimensional fields,
which include the graviton and any particles associated with the graviton by
supersymmetry. Cosmological equations, which describe the evolution from the
point of view of the brane, can be obtained by decomposing the higher
dimensional Einstein equations and applying boundary or junction conditions
for the brane \cite{shiromizu99}. These equations reduce to the standard four
dimensional equations for energies small compared to the vacuum energy scale
of the brane.

Our interest here is mainly in the effects of radiation generated by the
movement of the brane. We shall focus our attention on the radiation back
reaction force on the  brane. The radiation is closely related to the
radiation due to a moving mirror \cite{davies76}. The mirror, like the brane,
disturbs the vacuum fluctuations of the radiation field through the effect of
the boundary condition at the surface of the mirror. The vacuum fluctuations
react back on the mirror though the pressure component in the stress energy
tensor. 

It may be useful to recall some of the old results on moving mirrors with
massless scalar field radiation.  The  pressure force is proportional to the
3'rd time derivative of the mirror's position in two dimensions \cite{davies76}
and the 5'th time derivative in four dimensions\cite{ford82}. There is no
radiation reaction force on a uniformly accelerating mirror. A rather
perplexing result is that the pressure component of the stress energy tensor
diverges as the distance from the mirror shrinks to zero . It is believed that
the correct pressure on the mirror is obtained by dropping the divergent
terms. Some support for this view is provided by the fact that the total work
done by the divergent terms is zero, at least when the mirror motion has
constant velocity asymptotically.  

The moving mirror might also be placed in a heat bath with temperature $T$. In
this case, for a mirror moving with a speed  $v$, the thermal contribution to
the pressure force is proportional to $T^2v$ in two dimensions \cite{jaekel92}
and $T^4v$ in four dimensions \cite{machado02}. These agree with the forces
which would be expected according to elementary kinetic theory applied to a
mirror and a gas of photons.

The equation of motion for the mirror can be modified to take these radiation
reaction forces into account. The  modified equation of motion is analogous to
the Abraham-Lorentz equation of motion for an  electron with radiation
damping \cite{landau71}, which offers a guide to the interpretation of the
radiation reaction effects for the moving mirror and the moving brane. The
Abraham-Lorentz equation for the electron's momentum is
\begin{equation}
{dp\over dt}=F+\tau{d^2p\over dt^2} 
\end{equation}
where $F$ is an external force and $\tau=2e^2/3mc^3$. A feature of this
equation is the existence of  `runaway' solutions $p\propto \exp(t/\tau)$,
which are rejected on the grounds that the radiation reaction term should be
only a small correction to the equation of motion. The runaway solutions can
be eliminated by  replacing the equation by an integrodifferential equation, 
but this introduces another feature,  `preacceleration', where the electron
begins to respond a short time before the external force is applied. We shall
see that these features can be found in the brane system.

Recent work related to radiating branes includes \cite{setare04}, which
examined the radiation due to a uniformly accelerated brane. The authors did
not consider the radiation reaction,  but it would be expected to vanish for
this type of motion. The brane
can also radiate though interactions of the type $X+X\to Y$, where the $X$'s
are particles of matter and $Y$ is a bulk graviton. The effect of these
gravitons
on the brane persists, and resembles ordinary radiation.  In some very
specific models of brane cosmology, the gravitons can make a small
contribution to the expansion rate during the nucleosynthesis era. This has
been investigated by a number of authors
\cite{hebecker01,langlois02,langlois03}. A similar type of reaction, where $X$
is the inflaton and $Y$ a bulk scalar, could also change the nature of
reheating after a period of inflation \cite{enqvist04} .

The next section contains some general results for the radiation from moving
branes and mirrors. We take some care with the perturbation theory because we
would like to ensure that the calculations give the vacuum expectation value 
$\langle \hbox{in}|T_{ab}|\hbox{in}\rangle$ of the stress energy tensor 
rather than $\langle \hbox{out}|T_{ab}|\hbox{in}\rangle$. It is also important
to separate the radiation reaction calculation from effective action and
casimir energy calculations. From the point of view of the brane and its
cosmological evolution, the radiation back reaction effect is a dissipative
phenomenon and cannot be derived from a reduced effective action which only
depends on the degrees of freedom attached to the brane. However, we shall see
that the Schwinger-Keldeysh formalism gives a reduced effective action which
can be used to analyse both dissipative and non-dissipative effects. 

\section{Radiation back reaction}

\subsection{Backreaction equations}

Our approach to the back reaction on a brane moving in one extra dimension is
closely based on the back reaction problem for a moving mirror \cite{ford82}.
In both cases the radiation is produced because the radiation satisfies
boundary conditions on a moving boundary.  In moving mirror problems the back
reaction force is determined by the pressure components of the expectation
value of the stress energy tensor.

The back reaction of this radiation on the motion of a brane in five dimensions
can be obtained by reducing the five dimensional Einstein equations in the
manner described by Shiromizu, Maeda and Sasaki \cite{shiromizu99}. This
reduction assumes a reflection symmetry between to two sides of the brane. For
the $5$-dimensional Einstein equations we have
\begin{equation}
G_{ab}=\kappa_5^2T_{ab}
\end{equation}
For simplicity, we shall consider just a 5-dimensional cosmological constant
$\Lambda_5$, a brane vacuum energy $\lambda$ and 5-dimensional radiation,
\begin{equation}
T_{ab}=-\kappa_5^{-2}\Lambda_5 g_{ab}-\lambda h_{ab}\delta(\Sigma)+
\langle\hbox{in}|T^r_{ab}|\hbox{in}\rangle
\end{equation}
where $h_{ab}$ is the metric induced on the brane by the 5-dimensional metric
$g_{ab}$. Using the Gauss-Codacci equations,  Shiromizu, Maeda and Sasaki show
that the Einstein tensor ${}^{(4)}G_{\mu\nu}$  satisfies
\begin{equation}
{}^{(4)}G_{\mu\nu}+\Lambda h_{\mu\nu}={2\kappa_5^2\over 3}
\langle\hbox{in} |T^r_{\mu\nu}
+(T^r_{nn}-\frac14 T^r)h_{\mu\nu}|\hbox{in}\rangle
-E_{\mu\nu}\label{smk}
\end{equation}
where $E_{\mu\nu}$ is a projection of the Weyl tensor, $C_{n\mu n\nu}$ and
therefore trace-free. The 4-dimensional cosmological constant term and Newton's
constant are related to their 5-dimensional counterparts by
\begin{eqnarray}
\Lambda&=&{1\over 12}
\left( 6\Lambda_5+\kappa_5^4\lambda^2\right)\\
G&=&{\kappa_5^4\lambda\over 48\pi}\label{bigG}
\end{eqnarray}
In order to recover the correct low energy limit for the single brane we
require $\lambda>0$ and, for a small or vanishing cosmological constant,
$\Lambda_5<0$.

The $E_{\mu\nu}$ term can be eliminated by taking the trace of (\ref{smk}),
giving an equation for the Ricci scalar of the brane,
\begin{equation}
{}^{(4)}R=4\Lambda-2\kappa_5^2\langle\hbox{in}|T^r_{nn}|\hbox{in}\rangle
\label{rbr}
\end{equation}
In a cosmological context, this is the equation of motion of the scale factor.
The expectation value on the right
includes the radiation damping effect on the expansion rate of the universe. In
general, the expectation value depends on both the five dimensional metric and
the brane motion. However, in this paper we shall assume that the back
reaction of the radiation on the five dimensional metric can be neglected in
the quantum calculation.

For a spatially flat, homogeneous universe with scale factor $a(t)$ and
expansion rate $H(t)$, the cosmological evolution equation (\ref{rbr}) becomes
\begin{equation}
6\dot H+12H^2=4\Lambda 
-2\kappa_5^2\langle\hbox{in}|T^r_{nn}|\hbox{in}\rangle.\label{Heom}
\end{equation}
The initial condition as $t\to-\infty$ is provided by the Friedman equation
$3H^2=\Lambda$. The Friedman equation is modified at all other times by the
Weyl term,
\begin{equation}
E_{00}={2\kappa_5^2\over 3}
\langle\hbox{in}|(T^r_{00}-T^r_{nn})|\hbox{in}\rangle
+\Lambda-3H^2,
\end{equation}
which can be interpreted as energy lost by the brane. Note, however, that the
quantum term can be negative.

Typically,  the radiation reaction 
$\langle\hbox{in}|T^r_{nn}|\hbox{in}\rangle$ 
will only be significant for very early times, but it leaves behind a residual
`dark radiation' term \cite{binetruy99b} obtained by integrating (\ref{Heom}),
\begin{equation}
E_{00}\sim{2\kappa_5^2\over a^4}\int_{-\infty}^\infty
H(t')a(t')^4\langle\hbox{in}|T^r_{nn}(t')|\hbox{in}\rangle\,dt'\label{dr}
\end{equation}
In principle, this term behaves like a contribution to the radiation energy
density, but we have not, so far, considered the effects of ordinary matter.  
If, for example,  matter is generated by reheating after a period of
inflation, then the dark radiation produced during or prior to inflation would
be inflated away \cite{flanagan99}.

\subsection{Quantum stress energy tensor}

We shall obtain some general results for the expectation value of the stress
energy tensor using perturbation theory. The classical action of the radiation
field $\phi$ is
\begin{equation}
S_r=\frac12\int_{\cal M}\phi\Delta\phi\,dv\label{raction}
\end{equation}
where $\Delta$ is a second order operator and $dv$ is the volume measure. The
unperturbed system is chosen to
be in vacuum state where the green functions of the radiation with the
prescribed boundary conditions are known explicitly.  For scalar fields, the
boundary conditions can be Dirichlet, where the field vanishes on the
boundary, or Robin, where the normal derivatives of the field are prescribed.
Our notation for Green functions follows ref. \cite{birrell82}.

\begin{figure}
\scalebox{0.7}{\includegraphics{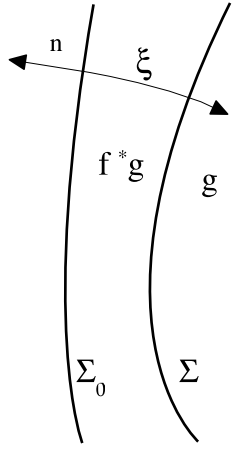}}
\caption{\label{fig1}The brane $\Sigma$ can be obtained by mapping the
unperturbed brane $\Sigma_0$ along the intergral curves of a vector field
$\xi$. The mapping is denoted by $f$ and the pullback mapping by $f^*$.}
\end{figure}

The relation between the perturbed and unperturbed boundaries can be defined by
a  diffeomorphism $f$ as shown in figure \ref{fig1}. If the perturbation is
small, then the vacuum state should be invariant. Expectation values
calculated for the  brane $\Sigma$ and the metric $g$ should be equivalent to
the expectation values calculated with the unperturbed brane $f^{-1}(\Sigma)$
and the  metric $f^* g$. By this means we can replace a perturbation of the
brane by a perturbation of the metric.

The number of dimensions  $n$ is left arbitrary.  Local coordinates can be
adapted to the unperturbed brane so that $n-1$ coordinates $x^\mu$ describe the
position on the brane and a coordinate $z=-x^n$ describes the distance to the
perturbed brane. The perturbed brane is then described by the function
$z(x^\mu)$.

 We begin with the Wightman functions $G^{\pm}$ of the scalar field $\phi$,
defined by
\begin{eqnarray}
G^+(x.x')&=&\langle\hbox{in}|\phi(x)\phi(x')|\hbox{in}\rangle\label{wight}\\
G^-(x.x')&=&\langle\hbox{in}|\phi(x')\phi(x)|\hbox{in}\rangle.
\end{eqnarray}
The Wightman functions satisfy the homogeneous wave equation
\begin{equation}
\Delta G^{\pm}=0
\end{equation}
Their sum defines the Hadamard function,
\begin{equation}
G^{(1)}=G^++G^-
\end{equation}
The change in Hadamard's function when the operator is perturbed by the moving
brane is 
\begin{equation}
\delta G^{(1)}=-G_R(\delta\Delta)G^{(1)}-G^{(1)}(\delta\Delta)G_A
\end{equation}
where $G_R$ and $G_A$ are the retarded and advanced propagators,
\begin{equation}
G_R(x,x')=G_A(x',x)=-i\langle\hbox{in}|[\phi(x)\phi(x')]|\hbox{in}\rangle
\theta(t-t')
\end{equation}
The perturbed form of the Hadamard function is real, symmetric and causal.

The first order change in the Hadamard function can be cast into a form which
is more convenient for calculations,
\begin{equation}
\delta G^{(1)}=-4{\rm Im}(G_0^>(\delta\Delta)G_0^<)
\end{equation}
where the subscript denotes the use of the unperturbed green function, and
\begin{equation}
G_0^>(x,x')=G_0^<(x',x)=iG_0^+(x,x')\theta(t-t')\label{gbig}
\end{equation}
The subscripts will subsequently be dropped. 

The next step is to find an expression for $\delta\Delta$. Consider the general
second order operator
\begin{equation}
\Delta=-\nabla^2+\xi R+m^2\label{operator}
\end{equation}
The change in the operator is induced by the coordinate transformation
$\xi^a=-z(x^\mu)n^a$ as shown in figure \ref{fig1}. The diffeomorphism
invariance of the classical action of the radiation (\ref{raction}) implies
that
\begin{equation}
\int_{\cal M}\left(\phi(\delta\Delta)\phi
+(\xi^a\nabla_a\phi)\Delta\phi+\phi\Delta(\xi^a\nabla_a\phi)
\right)dv+\int_{\Sigma}z\phi\Delta\phi\,dv=0
\end{equation}
After integration by parts we deduce that
\begin{equation}
\delta\Delta=\nabla_a\xi^a\Delta-\Delta\xi^a\nabla_a\label{ddelta}
\end{equation}
The required combination of green functions can now be found  explicitly by a
covariant volume integral,
\begin{equation}
\delta G^{(1)}=
4\,{\rm Im}\,\int_{\cal M}
G^>(x,x')(\Delta\xi^{a'}\nabla_{a'}-\nabla_{a'}\xi^{a'}\Delta)
G^>(y,x')dv'
\end{equation}
The imaginary part of the integrand consists of a total divergence which
reduces by the divergence theorem to
\begin{equation}
\delta G^{(1)}=
4\,{\rm Im}\,\int_{\Sigma}\left( \nabla_{n'}G^>(x,x')\nabla_{n'}G^>(y,x')
-G^>(x,x')\nabla_{n'}^2G^>(y,x')\right)z(x')dv'\label{deltag1}
\end{equation}

The reason for obtaining this expression for the Hadamard function is that it
can be used to find the expectation value  of the stress energy tensor. The
expectation value  of the stress energy tensor contains terms such as
\begin{equation}
\langle\hbox{in}|(\nabla_a\phi\nabla_b\phi
+\nabla_b\phi\nabla_a\phi)|\hbox{in}\rangle,
\end{equation}
which can be evaluated by applying an operator to the hadamard function and
taking a coincidence limit \cite{birrell82},
\begin{equation}
[\nabla_a\nabla_{b'}G^{(1)}].
\end{equation}
The brackets denote the coincidence limit $x=x'$. The full stress energy tensor
is given by
\begin{equation}
\langle\hbox{in}|T^r_{ab}|\hbox{in}\rangle=
\frac12[D_{a'b}G^{(1)}]\label{tdg}
\end{equation}
where the distribution valued operator $D_{ab}$  is given by
\begin{equation}
D_{ab}=\frac12{\delta^2\over \delta\phi^2} T_{ab}=
{\delta\Delta\over\delta g^{ab}}.\label{dab}
\end{equation}
In the minimal case, for example,
\begin{equation}
D_{ab}(x,x')=
\left(\nabla_{a'}\nabla_b-\frac12g_{ab}g^{cd}\nabla_{c'}\nabla_d\right)
\delta(y,x)\delta(y,x')
\end{equation}
For Dirichlet boundary conditions, the change in the normal component of the
stress energy tensor at the brane position is given by (\ref{deltag1}),
\begin{equation}
\delta\langle\hbox{in}|T^r_{nn}|\hbox{in}\rangle=
{\rm Im}\int_{\Sigma}
\left(\nabla_n\nabla_{n'}G^>(x,x')\right)^2z(x')\,dv'\label{tnn}
\end{equation}
which is equivalent to an old result due to Ford and Vilenkin \cite{ford82}.
For a moving
mirror, this stress energy component generates the pressure force on the
mirror. Since the unperturbed value of the stress energy does not affect the
motion we shall drop the $\delta$ and only keep the perturbed value from now
on.

\subsection{Schwinger-Keldeysh version}

Before proceeding, it is interesting to see how the same results can be
obtained more directly from an effective action using Schwinger-Keldeysh
methods \cite{schwinger61,keldysh64,calzetta87}. In this approach, the operator
$\Delta$ is a $2\times2$ matrix which
depends on two copies of the background fields,
\begin{equation}
\Delta_{SK}=\pmatrix{\Delta[g_1]&0\cr
0&\Delta[g_2]\cr}
\end{equation}
The radiation reaction forces can be obtained by variation
of the one loop  correction to the Schwinger-Keldeysh effective action,
\begin{equation}
W_{SK}=-\frac{i}2\log\det(iG_{SK})\label{actionsk}
\end{equation}
where $G_{SK}$ is the green function in the  Schwinger-Keldeysh approach. Let
$S[g_1]$ be the classical gravitational action, then 
$\delta_1 S+\delta_1 W_{SK}=0$,
where $\delta_1$ denotes variation of $g_1$, and $g=g_1=g_2$ after the
variation.

The Schwinger-Keldeysh green function is a $2\times2$ matrix
\begin{equation}
G_{SK}=\pmatrix{G_T&iG^-\cr-iG^+&-G_{\bar T}\cr}\label{skp}
\end{equation}
where
\begin{eqnarray}
G_T(x,x')&=&iG^+\theta(t-t')+iG^-\theta(t'-t)\\
G_{\bar T}(x.x')&=&iG^-\theta(t-t')+iG^+\theta(t'-t).
\end{eqnarray}
(Note that $G_T$ is identical to the Feynman Green function if the in and out
vacua are the same state, but not necessarily identical otherwise). The 
Schwinger-Keldeysh Green function changes under perturbations of the operator
by
\begin{equation}
\delta G_{SK}=-G_{SK}(\delta\Delta)_{SK}G_{SK}
\end{equation}
Multiplying out the matrices gives, for example,
\begin{equation}
\delta G_T=-G_T(\delta\Delta)G_T+G^-(\delta\Delta)G^+
\end{equation}
Under the same restrictions as before, this is equivalent to
\begin{equation}
\delta G_T=-2\,{\rm Im}(G^>(\delta\Delta)G^<)\label{pertsk}
\end{equation}
The variation of the Schwinger-Keldeysh effective action to first order in
perturbation theory is therefore
\begin{equation}
\delta_1W_{SK}={\rm Im}\,{\rm tr}((\delta_1\Delta)G^>(\delta\Delta)G^<).
\end{equation}
By (\ref{tdg}) and (\ref{dab}), this is equivalent to the previous result for
the stress energy tensor of the radiation. 

Alternatively, we we can regard the action as a function of the brane position
and obtain a reduced action formulation,  often called the moduli space
approximation. Let
\begin{equation}
W[z,g]=\frac12{\rm Im}\,{\rm tr}\,\left(\delta\Delta \,G^>\,\delta\Delta\,
G^<\right).
\end{equation}
Variation with respect to $z$ can be related to a metric variation $\delta_1W$
by diffeomorphism invariance, as in figure \ref{fig1}. The effective equations
of motion are then
\begin{equation}
{\delta S\over\delta z}+{\delta W\over\delta z}=0
\end{equation}

The Schwinger-Keldeysh formalism is usually applied to non-equilibrium thermal
field theory, using finite temperature version of the propagator (\ref{skp}).
Our results are therefore equally applicable to finite temperatures (for a
time-independent background) if we
replace the function $G^>(x,x')$ by the finite temperature version.

\section{Branes moving in flat space}

We shall examine the radiation back reaction force on the brane to leading
order in the displacement from a flat hyperplane in flat space. Consider the
massless field with Dirichlet boundary conditions to begin with.
The unperturbed green function (\ref{gbig}) can be expressed in terms of basis
functions which vanish at $z=0$,
\begin{equation}
G^>(x,x')=\int_0^\infty{dq\over 2\pi}\int{d^{n-2}k\over (2\pi)^{n-2}}
{i\over 2\omega}4\sin qz\,\sin qz'\,
e^{i{\bf k}({\bf x}-{\bf x}')-i\omega(t-t')}\theta(t-t')
\end{equation}
where $\omega=(k^2+q^2)^{1/2}$. Inserting this into equation (\ref{tnn}) for
the stress energy tensor gives
\begin{equation}
\langle\hbox{in}|T^r_{nn}|\hbox{in}\rangle=
{\rm Im}\,{\rm reg}\int_{-\infty}^\infty{dq\over 2\pi}{dq'\over 2\pi}
\int{d^{n-2}k\over (2\pi)^{n-2}}{q^2q^{\prime 2}\over \omega\omega'}
I[z]
\end{equation}
where `reg' indicates some form of regularisation has been performed to make
the integral finite and
\begin{equation}
I[z]=\int_{-\infty}^te^{-i(\omega+\omega')(t-t')}z(t')dt'
=-i\left((\omega+\omega')-i\partial_t\right)^{-1}z
\end{equation}
Note the importance of taking the imaginary part of the expression {\it after}
regularisation. It is useful to introduce
\begin{equation}
F_n(x)={\rm reg}\,\int{dq\over 2\pi}{dq'\over 2\pi}
{d^{n-2}k\over (2\pi)^{n-2}}{q^2q^{\prime 2}\over \omega\omega'}
{-i\over (\omega+\omega')-ix}
\end{equation}
and then
\begin{equation}
\langle\hbox{in}|T^r_{nn}|\hbox{in}\rangle=
{\rm Im}\, F_n(\partial_t)z
\end{equation}

The integral diverges for all values of the dimension $n$ which rules out
dimensional regularisation. However, an analytic regularisation scheme can be
used where we define
\begin{equation}
F_{ns}(x)={i\over \Gamma(s+1)}
\int_0^\infty d\lambda \lambda^s\int{dq\over 2\pi}{dq'\over 2\pi}
{d^{n-2}k\over (2\pi)^{n-2}}{-q^2q^{\prime 2}\over
\omega\omega'}e^{-\lambda(\omega+\omega')+i\lambda x}\label{fns}
\end{equation}
and take the value at $s=0$, removing pole terms if necessary. After
integrating over $q$, we have
\begin{equation}
F_{ns}(x)={-4i\over \Gamma(s+1)}
\int_0^\infty d\lambda \lambda^s\int
{d^{n-2}k\over (2\pi)^{n-2}} (\lambda k)^{-2}K_1(\lambda k)^2\,e^{i\lambda x}
\end{equation}
where $K_1$ is a Bessel function of the second kind. The remaining integrals
give
\begin{equation}
F_{ns}(x)={1\over\Gamma(s+1)}{x^{n-s+1}\over 2\pi^{n/2}}
{\Gamma(n/2)^3\over (n+1)\Gamma(n)^2}{i^{n-s}\over \sin\pi(n-s)}
\end{equation}
After removing the pole at $s=0$, the regularised expression is
\begin{equation}
F_n(x)=-{(-ix)^{n+1}\over 4\pi^{n/2}}
\left(1-i\ln\left({x\over\mu}\right)\right)
{\Gamma(n/2)^3\over(n+1)\Gamma(n)^2}
\end{equation}
where $\mu$ is a renormalisation constant.

Some examples are
\begin{eqnarray}
{\rm Im}\,F_2(x)&=&-{x^3\over 12\pi}\\
{\rm Im}\,F_4(x)&=&{x^5\over 720\pi^2}\\
{\rm Im}\,F_5(x)&=&-{x^6\over16384\pi^2}\ln\left({x\over\mu}\right)
\end{eqnarray}
The results for two and four dimensions agree with those found by Ford and
Vilenkin for the moving mirror problem\cite{ford82}. 

The dependence on $\mu$ in odd dimensions can be traced to need for new
counterterms in the classical action \cite{moss03} which arise from the $n$'th
heat kernel coefficient of the operator $\Delta$ \cite{gibbons03}. In five
dimensions, the heat kernel coefficient contains boundary terms of the form
\begin{equation}
S_{5s}={1\over s}\int_\Sigma\,\nabla_\mu K_{\nu\rho}
\nabla^\mu K^{\nu\rho}\,dv,
\end{equation}
and other permutations of the indices, where $K$ is the extrinsic curvature of
the brane. These  counterterms produce a ${\partial_t}^6z$ in the equation of
motion (since $K=-{\partial_t}^2z$ to leading order) which cancels the
divergence in the radiation reaction and leaves the logarithmic dependence on
$\mu$.  How the logarithmic derivatives  are interpreted, and how to solve the
equations of motion for the brane will be covered in section 4.

The radiation back reaction from a massive scalar field with mass $m$ can be
found by redefining  $\omega=(k^2+q^2+m^2)^{1/2}$ in eq. (\ref{fns}). In
even dimensions, the integral results in
\begin{equation}
{\rm Im}\,F_n(x)={1\over 8(4\pi)^{n/2}}m^{n-1}x^2
\Gamma\left(\frac{1-n}2\right)\Gamma\left(\frac12\right)
{}_2F_1\left(\frac{1-n}2,\frac12;2;-\frac{x^2}{4m^2}\right)
\end{equation}
where ${}_2F_1$ is a hypergeometric function. In odd dimensions,
we discard the pole term and retain only the finite part of the same
expression. 

For small values of $x$, $F_n$ is of order $x^2$.  If the position of the brane
is oscillating with a frequency $\omega<m$, the radiation reaction force is
proportional to $\omega^2$, which is similar to a known result for a domain
wall in four dimensions \cite{vachaspati84}. For large values of $x$, the
hypergeometric functions have
branch cuts in the region $|x|>2m$ and logarithmic terms appear in the large
$x$ limits for both even and odd dimensions. The leading terms agree with the
massless results given above.

\section{Branes moving in anti-De Sitter space}

Branes in anti-de Sitter space are interesting from a cosmological point of
view. The intrinsic geometry of a homogeneous brane moving in anti-de Sitter
space is similar to a cosmological model \cite{kraus99,kehagias99}.  The
effects of the higher dimensional cosmological constant and the vacuum energy
on the brane can be fine-tuned to give a relatively small effective
cosmological constant on the brane \cite{randall99a,randall99b}. 

\begin{figure}
\scalebox{0.7}{\includegraphics{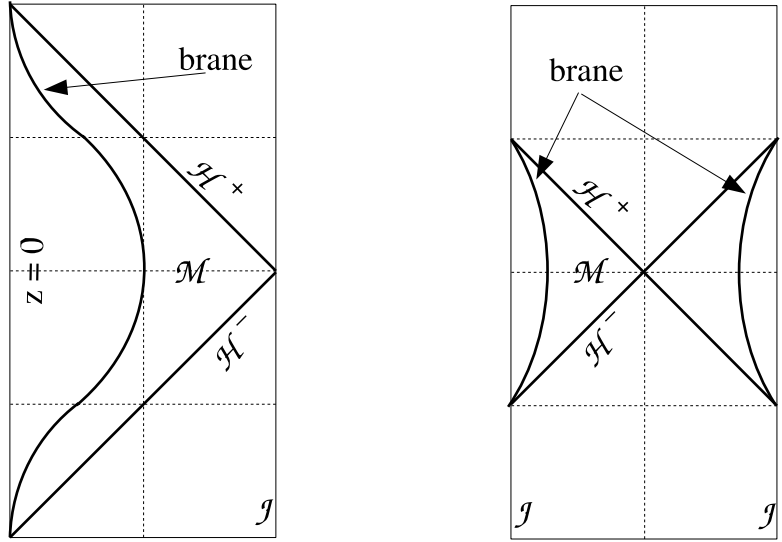}}
\caption{\label{fig2} On the left is a Penrose diagram of universal anti-de
Sitter space showing the location of a timelike slice of the flat brane. The
coordinate system extends on the right of the brane ($z>\sigma$) to the
horizons ${\cal H}^+$ and ${\cal H}^-$ where the timelike killing vector
becomes null. On the right is a copy of the Penrose diagram to show the
location of a de Sitter space brane.}
\end{figure}

As before, we will neglect the back reaction of the motion of the brane on the
bulk metric when calculating the quantum stress energy tensor. We shall
consider small perturbations of a single flat brane whose penrose diagram is
shown in figure \ref{fig2}. We take the vacuum state related to the timelike
translation symmetry along the brane, which we call the brane vacuum.

The brane vacuum has the disadvantage that there are horizons where the
timelike killing field vanishes. The horizons can be avoided, for example by
considering a two brane system and taking the limit where the separation
between the branes becomes large. If, instead, we use a vacuum state which
covers the whole of anti-de Sitter space, then we have to consider the effects
of Bogolubov coefficients \cite{birrell82}. In the case of the de Sitter
brane, the Euclidean vacuum state can be used \cite{moss03}. The Euclidean
vacuum becomes a thermal state with respect to the brane, creating additional
thermal effects in the radiation reaction force. We shall not consider this
further here.

One of the many ways to express the Anti-de Sitter metric in $n$ dimensions is
the conformally flat form,
\begin{equation}
\sigma^2z^{-2}(dz^2+\eta_{\mu\nu}dx^\mu dx^\nu)
\end{equation}
with Minkowski metric $\eta_{\mu\nu}$. The Anti-de Sitter radius $\sigma$ is
related to the cosmological constant in $n$ dimensions by
\begin{equation}
\Lambda_n=-{(n-1)(n-2)\over2\sigma^2}. 
\end{equation}
The  unperturbed brane will  be placed at $z=\sigma$ and the manifold
extends in the $z>\sigma$ direction only. The covering space can be filled in
by reflection symmetry about the brane.

The Green functions can be obtained from the normalised modes $u_{kq}$ of the
scalar wave equation. These are
\begin{equation}
u_{kq}={1\over (2\omega)^{1/2}}
\left({z\over\sigma}\right)^{(n-1)/2}
H_\nu(q,z)e^{i({\bf k}\cdot{\bf x}-\omega t)}
\end{equation}
with $\omega=(k^2+q^2)^{1/2}$ and $H_\nu$ a combination of Hankel functions
of order
\begin{equation}
\nu^2=\frac14+\sigma^2\left((\xi-\xi(n))R+m^2\right).
\end{equation}
Conformal curvature coupling corresponds to the case $\xi=\xi(n)$ and
$\nu=1/2$.

For modes which vanish on the brane,
\begin{equation}
H_\nu(q,z)={i(\pi q\sigma)^{1/2}\over 2|H^{(1)}_\nu(q\sigma)|}
\left(H_\nu^{(1)}(qz)H_\nu^{(2)}(q\sigma)
-H_\nu^{(2)}(qz)H_\nu^{(1)}(q\sigma)\right)
\end{equation}
Note that the normal derivatives of the mode functions when evaluated on the
brane $z=\sigma$  are  then given by
\begin{equation}
\nabla_n u_{kq}={1\over (2\omega)^{1/2}}
\left({2q\sigma\over \pi}\right)^{1/2}{2^{1/2}q\over|H^{(1)}_\nu(q\sigma)|}
\,e^{i({\bf k}\cdot{\bf x}-\omega t)}
\end{equation}
In the conformal case $\nu=1/2$, this reduces to
\begin{equation}
\nabla_n u_{kq}={1\over (2\omega)^{1/2}}
\,2^{1/2}q\, e^{i({\bf k}\cdot{\bf x}-\omega t)}
\end{equation}
which is also the value obtained in flat space.

The normal derivatives of the Wightman function are needed to obtain the stress
energy tensor. The Wightman function can be obtained from a mode sum,
\begin{equation}
G^+(x,x')=\int_0^\infty{dq\over2\pi}\int{d^{n-2} k\over (2\pi)^{n-2}}
\,u_{kq}(x)u^*_{kq}(x')
\end{equation}
The normal derivatives at the brane can be expressed as
\begin{equation}
\nabla_n\nabla_{n'}G^+(x,x')=\int_0^\infty{dq\over2\pi}
\int{d^{n-2} k\over (2\pi)^{n-2}} 
{2q^2\over 2\omega} \left|{H^{(1)}_{1/2}(q\sigma)\over
H^{(1)}_\nu(q\sigma)}\right|^2
 e^{i({\bf k}\cdot({\bf x}-{\bf x'})-\omega (t-t'))}
\end{equation}
As in the flat space example, the expectation value of the normal components of
the stress energy tensor are given in terms of an integral by
\begin{equation}
\langle\hbox{in}|T^r_{nn}|\hbox{in}\rangle=
{\rm Im}\, F_n(\partial_t)z
\end{equation}
where the anti-De Sitter space version of $F$ is
\begin{equation}
F_n(x)={\rm reg}\,\int{dq\over 2\pi}{dq'\over 2\pi}
{d^{n-2}k\over (2\pi)^{n-2}}{q^2q^{\prime 2}\over \omega\omega'}
\left|{H^{(1)}_{1/2}(q\sigma)H^{(1)}_{1/2}(q'\sigma)\over
H^{(1)}_\nu(q\sigma)H^{(1)}_\nu(q'\sigma)}\right|^2
{-i\over (\omega+\omega')-ix}\label{newf}
\end{equation}

The first striking feature of eq. (\ref{newf}) is that,  in the conformal case
$\nu=1/2$, the result for radiative back reaction is identical to the flat
space result. In the non-conformal case, we can obtain the large $\sigma x$
limit from the large argument expansion of the Hankel functions, for example
\begin{equation}
{\rm Im}\,F_5(x)\sim -{x^6\over16384\pi^2}\ln\left({x\over\mu}\right)
\label{F5}
\end{equation}
in five dimensions. The small $\sigma x$ limit can correspondingly be obtained
from the small argument expansion of the Hankel functions, which leads to
\begin{equation}
{\rm Im}\,F_5(x)\sim C x^6(\sigma x)^{4\nu-2}
\end{equation}
when $\sigma x\ll 1$, where $C$ is a constant.

\section{Equations of motion with logarithmic terms}

We can now construct the equation of motion for the brane. For a small
perturbation $z$ of a flat brane, the scale factor $a\approx1$ and
$\partial_t z\approx-\sigma H$. For conformal scalar field radiation, eq.
(\ref{F5}) suggests that eq. (\ref{Heom}) would become
\begin{equation}
{dH\over dt}+2H^2-2H_0^2=
-A\kappa_5^2\sigma\ln\left({1\over\mu}{d\over dt}\right){d^5H\over dt^5}.
\label{neweom}
\end{equation}
where $H_0^2=\Lambda/3$ and $A$ is a numerical coefficient. Strictly speaking,
this equation is incomplete because there may be additional radiation damping
terms of order $H^2$.

The radiation damping term calculated here is small in recent cosmological
eras. Returning to eq (\ref{bigG}), we see that the
combination $\kappa_5^2\sigma\sim G\sigma^2$ in order of magnitude. The
experimental lower bound on $(G\sigma^2)^{-1/4}$ is around $1TeV$, but there
is no reason for the the value to be much smaller than the usual Planck scale.
When $\dot H\sim H^2$, the damping term is only important when $H^4\sim 
(G\sigma^2)^{-1}$. 

There is a small residual dark radiation effect given by
eq.  (\ref{dr}).  If we take into account a period of inflation, ending at
time $t_I$, then the ratio of dark radiation to ordinary radiation is
\begin{equation}
6A\kappa_5^2\sigma\int_{t_I}^\infty{H\over \rho}
\ln\left({1\over\mu}{d\over dt}\right){d^5H\over dt^5}dt\label{dri}
\end{equation}
which is of order $G\sigma^2H(t_I)^4$. In principle, $H(t_I)$ should be close
to the value of $H$ determined by the  tensor perturbation amplitude $A_T$ in
the cosmic microwave background,  $A_T^2=8\pi GH^2$ \cite{liddle93}. To be
within nucleosynthesis constraints, the amount of dark radiation must be small
\cite{flanagan99,hebecker01}, and therefore there is a lower bound on the
string vacuum energy scale of approximately
\begin{equation}
(G\sigma^2)^{-1/4}>A_T^{-1/2}m_p
\end{equation}
where $m_p$ is the Planck mass.

We shall analyse a slightly more general type of equation of motion,
\begin{equation}
{d p\over dt}-F(p)=-\eta
\ln\left({1\over\mu}{d\over dt}\right){d^np\over dt^n}\label{motion}
\end{equation}
A reasonable definition of the logarithmic term should be linear and causal, to
ensure that the back reaction depends on the history of the source. We can
begin be defining the action of the logarithmic derivative on exponentials,
\begin{equation}
\ln\left({1\over\mu}{d\over dt}\right)\,e^{\alpha t}
=\ln\left({\alpha\over\mu}\right)\,e^{\alpha t}
\end{equation}
where we take $\alpha>0$. This leads to the following definition,
\begin{equation}
\ln\left({1\over\mu}{d\over dt}\right)f
=-\int_{-\infty}^t\ln\left(e^\gamma\mu(t-t')\right){df\over dt'}dt'
\end{equation}
where $\gamma$ is Euler's constant. The definition is linear, causal and
correctly reproduces the action on exponentials. The integral can be evaluated
for a broad class of functions, including discontinuous functions, for example
\begin{equation}
\ln\left({1\over\mu}{d\over dt}\right)\theta(t)e^{\alpha t}=
\left({\rm E}_1(\alpha t)+\ln\left({\alpha\over\mu}\right)
\right)\theta(t)e^{\alpha t}
\end{equation}
where ${\rm E}_1$ is the exponential integral. If $f$ is integrable, then the
logarithmic derivative of $f$ is $O(t^{-1})$ as $t\to\infty$.

\begin{figure}
\scalebox{1.0}{\includegraphics{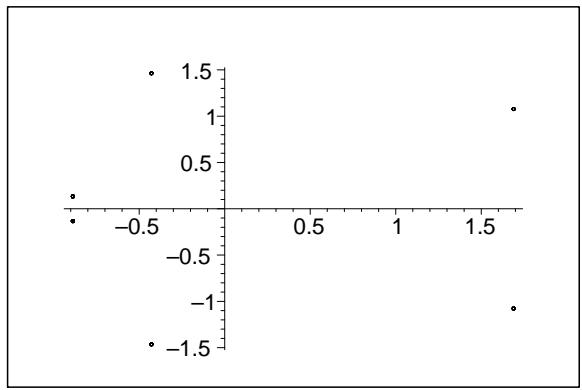}}
\caption{\label{fig3} The roots of  $1+z=-0.1z^5\ln(z)$ in the complex plane.}
\end{figure}

For a linear equation of motion (\ref{motion}), with $F=-\lambda p$, we have
\begin{equation}
 {d p\over dt}+\lambda p=-\eta
\ln\left({1\over\mu}{d\over dt}\right){d^5p\over dt^5}
\end{equation}
The exponential solutions $\exp(\mu zt)$ satisfy
\begin{equation}
z+\alpha=-\beta z^n\ln(z)
\end{equation}
where $\alpha=\mu^{-1}\lambda$ and  $\beta=\eta\mu^{n-1}$.
The roots for a sample case are plotted in figure \ref{fig3}. When there is
no reaction term there is only a decaying mode corresponding to $z=-\alpha$.
The reaction term shifts this solution and introduces an oscillation. There
are two other decaying solutions and two growing modes. These modes have 
$|z|>1$ and damping or growth rates larger than the renormalisation scale
$\mu$. We tentatively identify these as unphysical runaway solutions.

Runaway solutions can be excluded by replacing the equation of motion by  an
integrodifferential equation. First, we introduce a Green function which
decays exponentially as the time $t\to\pm\infty$ and satisfies the following
equation,
\begin{equation}
G+\eta
\ln\left({1\over\mu}{d\over dt}\right){d^{n-1}G\over dt^{n-1}}=\delta(t-t')
\end{equation}
The equation of motion (\ref{motion}) can then be rewritten in an alternative
form
\begin{equation}
{dp\over dt}=\int_{-\infty}^\infty G(t-t')F(t').
\end{equation}
The runaway solutions are excluded at the expense of a preacceleration term, as
we can see from the expression for the Green function
\begin{equation}
G(t-t')=
\sum_{{\rm Re}z_i<0}c_ie^{z_i (t-t')}\theta(t-t')
-\sum_{{\rm Re}z_i>0}c_ie^{z_i (t-t')}\theta(t'-t),
\end{equation}
where $z_i$ are the roots of $\eta z^{n-1}\ln(z/\mu)=-1$ and $c_i$ are
constants fixed by continuity relations at $t=t'$. The acausality is
represented by the final term, but it is a
relatively minor effect if ${\rm Re}z_i\gg\mu$, because then the acausality
only occurs on timescales small compared to the renormalisation timescale.

An interesting variant of the equation of motion (\ref{motion}) occurs for the
case $n=1$ and $F=0$. There are two real solutions,  $p=0$ and
$p=\exp(\exp(-\mu\eta^{-1})t)$.  Because of the small size of the exponent it
would not be appropriate to regard the exponential solution as a runaway
solution. In a cosmological context, a flat brane would
spontaneously begin to accelerate with a tiny, positive cosmological constant.
We have not been able to find a model with this particular radiation reaction
so far.

\section{Conclusion}

This paper has been devoted to calculating the radiation back reaction forces
on a moving brane. We have found that massless, conformally invariant scalar
radiation into the bulk results in logarithmic terms in the equations of
motion. The radiation reaction leads to both physical and runaway
solutions. The runaway solutions can be  excluded by introducing an
integrodifferential formulation of the equation of motion. This form of the
equation could be used if we were to attempt to solve the equations of motion
numerically. The question of whether the  brane's initial trajectory might be
destabilised by radiation reaction forces remains an open one.

The radiation from the moving brane also leaves a residual effect in the form
of dark radiation. The amount of this radiation depends on the brane
trajectory (\ref{dri}). An order of magnitude estimate combined with a crude
nucleosynthesis constaint gives a lower bound on the string vacuum energy
scale of $\sqrt{A_T}\,m_p$, where $A_T$ is the tensor perturbation amplitude in
the cosmic microwave background.

The radiation reaction problem has been set up in a way which can be
generalised to non-scalar fields. One issue which would be of interest is the
possibility of supersymetric cancellations amongst some of the terms. The
regularisation methods used in this paper have enabled us to neglect the
non-logarithmic divergences, but it would be better to see direct cancellation
of these divergent terms. For broken supersymmetry, we expect that the
renormalisation scale, which appears in our results, would be replaced by a
supersymmetry breaking scale. 

The results for the radiation reaction of branes moving in anti-de Sitter space
generalises in a very staightforward way to any background where the modes of
the radiation fields are known explicitly. An example of this is the low
energy limit of the heterotic string \cite{horava96-2,lukas98,lukas98-2},
where the modes of the graviton multiplet are known and can be found in
\cite{moss04}.

Another aspect of the radiation reaction which is worth further study is the
effect of different higher dimensional vacuum states. These states can have
thermal properties and the reaction forces would have similarities to those
acting on moving mirrors at finite temperatures. 

%%%%%%%%%%%%%%%%%%%%%%%%%%%%%%%%%%%%%%%%%%%%%
\bibliography{radiation.bib}

%%%%%%%%%%%%%%%%%%%%%%%%%%%%%%%%%%%%%%%%%%%%%%
\end{document}